\documentclass[preprint,prb,showpacs,amsmath,amssymb,12pt]{revtex4}
\usepackage{color,graphicx}
\usepackage{bm}
\begin{document}

\title{Anisotropic In-Plane Strain and Transport in Epitaxial
Nd$_{0.2}$Sr$_{0.8}$MnO$_3$ Thin Films}
\author{K. P. Neupane,$^{1\dagger}$ J. J. Neumeier,$^2$ and J. L. Cohn$^1$}
\affiliation{$^{1}$ Department of Physics, University of Miami, Coral Gables, Florida 33124}
\affiliation{$^2$ Department of Physics, Montana State University, Bozeman, Montana 59717}

\begin{abstract}
The structure, morphology, and electrical properties of epitaxial
$a$-axis oriented thin films of Nd$_{0.2}$Sr$_{0.8}$MnO$_3$ are
reported for thicknesses $10\ {\rm nm}\leq t \leq 150$~nm.
Films were grown with both tensile and compressive strain on various substrates.
It is found that the elongated crystallographic $c$-axes of the films remain fully strained
to the substrates for all thicknesses in both strain states.
Relaxation of the $a$ and $b$ axes is observed for $t\gtrsim 65$~nm,
with films grown under tensile strain developing uniaxial crack arrays
(running along the $c$ axis) due to a highly anisotropic thermal expansion.
For the latter films, the room-temperature in-plane electrical resistivity
anisotropy, $\rho_b/\rho_c$, increases approximately exponentially
with increasing film thickness to values of $\sim 1000$ in the
thickest films studied. Films under tension have their N\'eel temperatures enhanced
by $\approx 25$~K independent of thickness, consistent with an enhancement of ferromagnetic
exchange along their expanded $c$ axes.

\end{abstract}

\maketitle

\section{\label{sec:Intro} INTRODUCTION}

Using strain in thin films to achieve expanded or contracted
lattices of novel materials has proven fruitful in producing new
phases or functionalities for known bulk compounds. An interesting
avenue of investigation is the growth of anisotropically strained
films of perovskite oxides. The heavily hole-doped perovskite
manganites that order in the C-type antiferromagnetic (AF)
arrangement,\cite{Kajimoto,Tobe} e.g. Nd$_{1-x}$Sr$_{x}$MnO$_3$ with $0.75\leq x\leq 0.9$,
are of interest in this regard. Single crystals of compounds with $x\gtrsim 0.8$ have not been
successfully grown to our knowledge.  The C-type AF state consists of
a one-dimensional ordering of $d_{3z^2-r^2}$ orbitals of the Mn$^{3+}$ sites
oriented along the elongated $c$ axis of the tetragonal structure, facilitating ferromagnetic (FM)
coupling of Mn$^{3+}$ and Mn$^{4+}$ spins along $c$ and an AF spin alignment
along the $a$ axes.\cite{CAFExpt}  $a$-axis oriented films under anisotropic strain should allow
for the manipulation of the FM double-exchange and AF superexchange interactions responsible for this
ordering.\cite{CAFTheory,CAFStability}
They should make accessible the study of intrinsic electrical anisotropy, a motivation for which is
the large static dielectric constant ($\epsilon\sim 100$) observed\cite{CohnEps} in polycrystalline
Ca$_{0.2}$La$_{0.8}$MnO$_3$, and hypothesized to arise from an enhanced polarizability
along the one-dimensional $c$-axis in the AF phase.

Relevant to the successful growth of $a$-axis films of these compounds
are thermal expansion coefficients along the $a$ and $c$ axes that are
opposite in sign in a broad temperature range
from the cubic-tetragonal transition to well below room
temperature (a consequence of the orbital ordering).\cite{Tobe}
This differential thermal expansion induces highly anisotropic and
temperature-dependent in-plane strain.  We find
that thicker films under anisotropic tensile strain within the substrate plane
develop uniaxial crack arrays with
regular crack spacing on the order of 1 $\mu$m.  These crack
arrays yield in-plane electrical resistivities that are highly
anisotropic with an anisotropy ratio that varies approximately
exponentially with film thickness, reaching values $\gtrsim 10^3$.
The N\'eel temperatures of films under tensile strain are enhanced by 25~K, independent of thickness,
over those of the bulk target and compressively strained films,
consistent with an increased stability of the orbital and spin order that is controlled
principally by the $c$-axis length.

\section{\label{sec:Expt} EXPERIMENT}

The polycrystalline target of nominal composition Nd$_{0.2}$Sr$_{0.8}$MnO$_3$ (NSMO)
was prepared by conventional solid state reaction as described elsewhere.~\cite{Terashita}
Thin films were grown by pulsed laser deposition on pseudocubic
(110)-oriented (LaAlO$_3$)$_{0.3}$-(Sr$_2$AlTaO$_6$)$_{0.7}$ (LSAT; $a=0.3868$~nm) and LaAlO$_3$ (LAO; $a=0.379$~nm),
and orthorhombic (100)-oriented NdGaO$_3$ (NGO; $a=0.543$~nm, $b=0.550$~nm, $c=0.771$~nm)
substrates. An excimer laser (KrF:$\lambda=248$~nm)
with a frequency of 10~Hz and and an energy density  $\sim$1-2~J/cm$^{2}$ were employed.  The substrate temperature
was $750^{\circ}$C and partial oxygen pressure 260~mTorr.
Following deposition, films were cooled in $\sim 760$ Torr O$_2$ at
$1^{\circ}$C/min. to $500^{\circ}$C and held for 1 hour
before cooling to room temperature.  The crystallographic orientation, film thickness~($t$) and
lattice constants were evaluated using a Philips X'Pert x-ray
diffractometer (Cu K$_{\alpha}$ radiation). Surface morphology was studied
with scanning electron microscopy (SEM).  Four-probe, in-plane dc resistivity (with silver epoxy contacts)
was measured on specimens with typical dimensions $6\times
1.3 \times t$~mm$^3$.  The room-temperature thermopower was measured for all specimens
with a steady-state method using gold leads and a chromel-constantan thermocouple.

The target lattice constants, $a=0.5390(5)$~nm, $c=0.766(1)$~nm are in reasonable agreement with
those reported by Kajimoto {\it et al.}\cite{Kajimoto} for crushed, melt-grown crystals at lower doping.
The value of the N\'eel temperature, $T_N\simeq 242$~K was inferred from the peak in $d\log\rho/d(1/T)$ (further discussed below).

\section{\label{sec:Disc} RESULTS AND DISCUSSION}

\subsection{\label{Structure} Lattice Constants and Morphology}

\begin{figure}
\includegraphics[width=3.375in,clip]{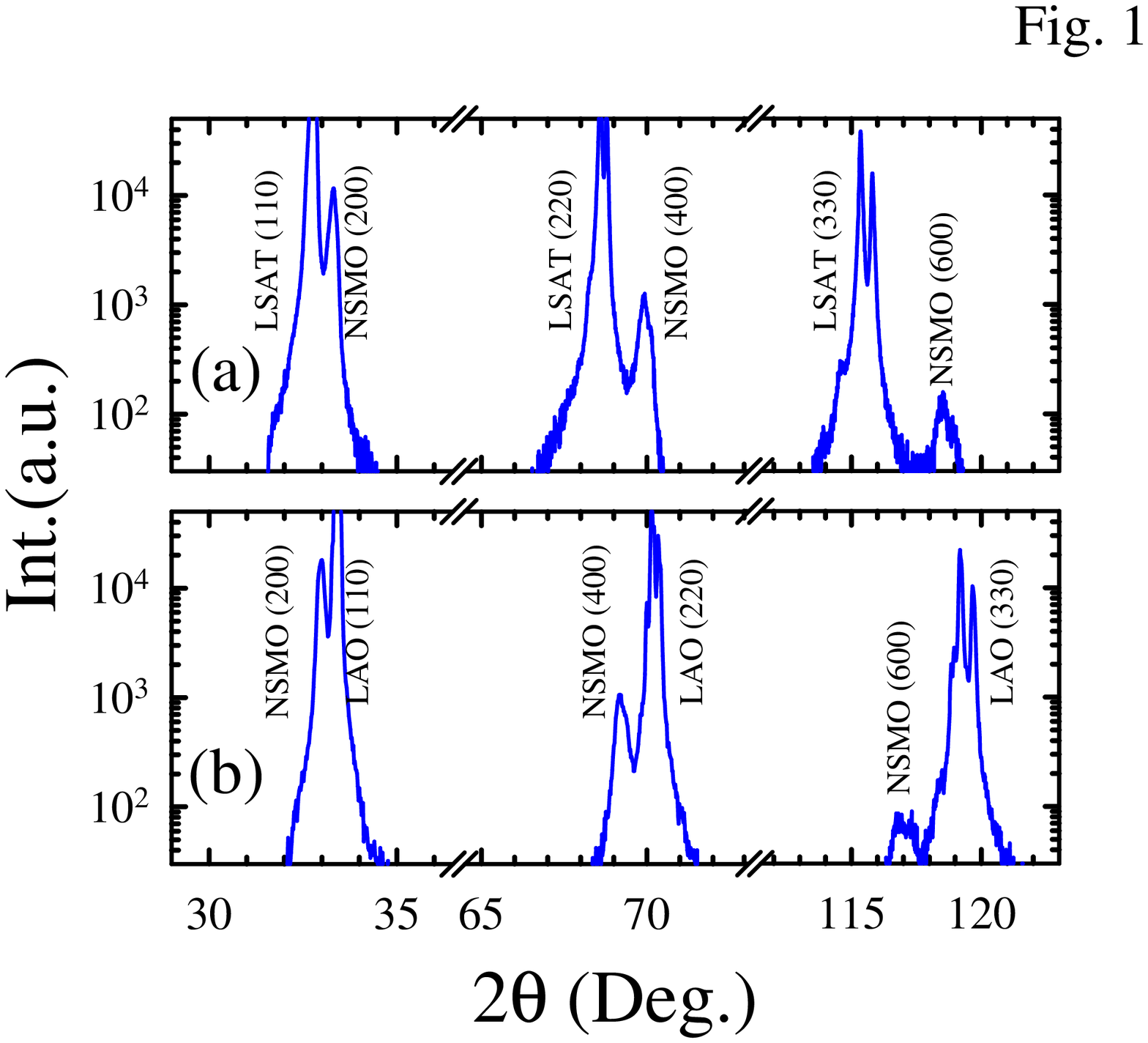}
\caption{(color online)$2\theta/\omega$ scans of NSMO films: (a) 130-nm on (110) LSAT, (b) 150-nm on (110) LAO.}
 \label{XRD}
\end{figure}

\begin{figure}
\includegraphics[width=3.375in,clip]{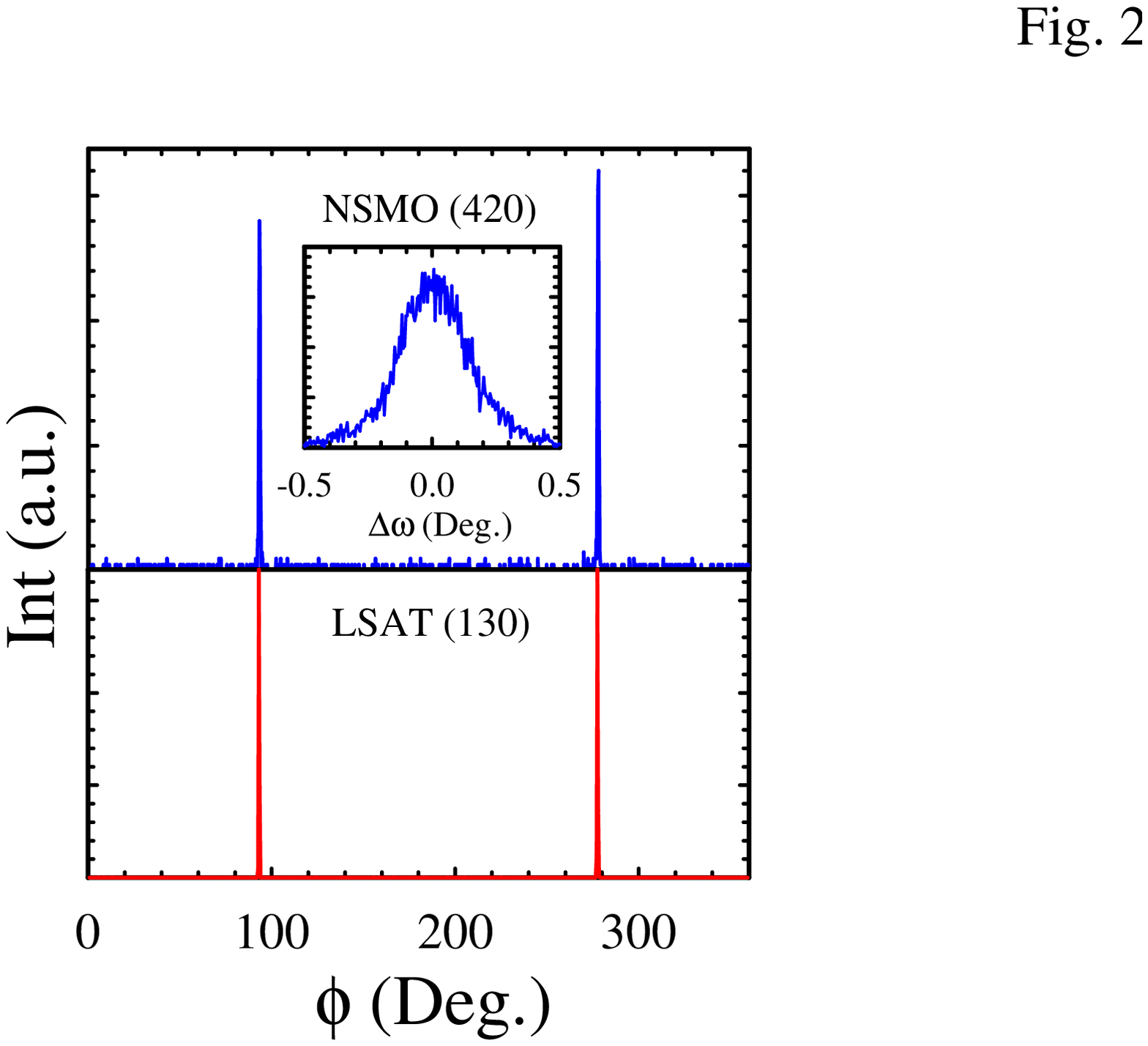}
\caption{(color online) Azimuthal ($\phi$) scans of the (420) film (upper panel) and (130) substrate (lower panel) reflections
for the 130-nm film grown on (110) LSAT, confirming epitaxial growth.  The inset shows the rocking-curve width for the film reflection.}
 \label{PhiScans}
\end{figure}

All of the films are orthorhombic with their longer $c$ axes in
the film plane. Taking the $a$ axis in the growth direction, the
(100) orientation of the films is indicated in x-ray diffraction
$2\theta$-$\omega$ scans, shown for 130 nm-thick films grown on
LSAT (110) and LAO (110) in Fig.~\ref{XRD} (a) and (b),
respectively, by the presence of only $(2h,0,0)$ film reflections
near the $(h,h,0)$ substrate reflections.  Phi scans of asymmetric
film and substrate reflections, shown in Fig.~\ref{PhiScans} for
the 130-nm film grown on LSAT, confirm the cube-on-cube
orientation with NSMO [010]$\parallel$LSAT [1$\overline{1}$0] and
NSMO [001]$\parallel$LSAT [001]; the same orientation relationship
was found for films grown on LAO.  X-ray results for the films on
NGO indicate NSMO [010]$\parallel$NGO [010] and NSMO
[001]$\parallel$NGO [001].

\begin{figure}
\includegraphics[width=3.375in,clip]{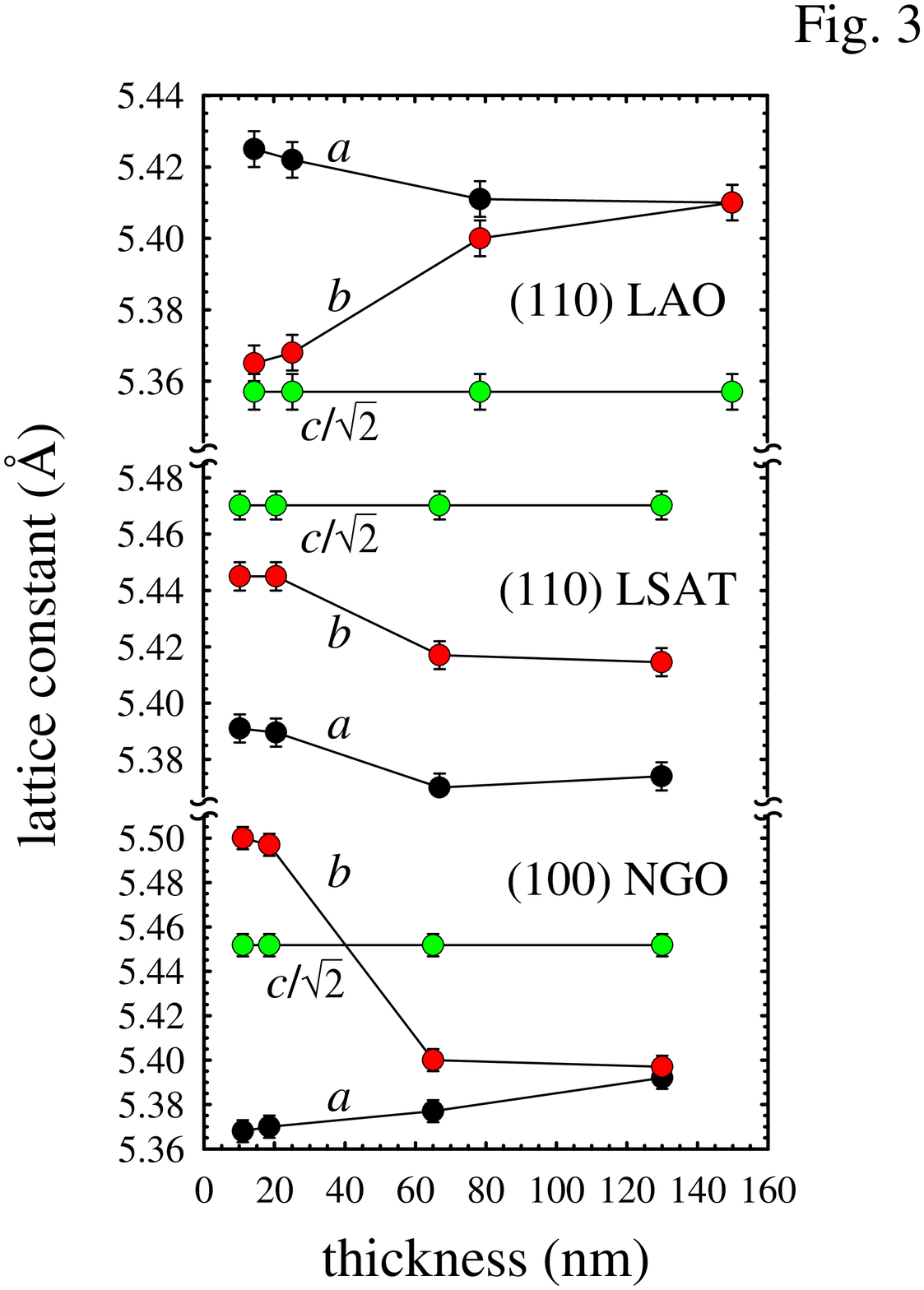}
\caption{(color online) Lattice constants {\it vs} thickness for NSMO films on three substrates.}
 \label{LatticeConstants}
\end{figure}

The lattice constants for all films were determined from reciprocal space maps in the vicinity of the
the film (600), (440), (620), and (404) reflections, with nearby substrate reflections serving as internal references.
The corresponding NGO reflections have the same indices as the films.  For (110)-oriented LAO and LSAT the corresponding substrate
reflections are (330), (400), (420), and (222), respectively.  Figure~\ref{LatticeConstants} shows lattice constants
as a function of film thickness for films on the three substrates.  For all substrates and thicknesses, the
film $c$-axes are fully strained to those of the substrate.  The $a$ and $b$ axis lengths are clearly relaxed in response to
the compressive (LAO) and tensile (LSAT, NGO) in-plane strain along the film [010] directions, with the most substantial effect
occurring for NGO.

\begin{figure}
\includegraphics[width=3.375in,clip]{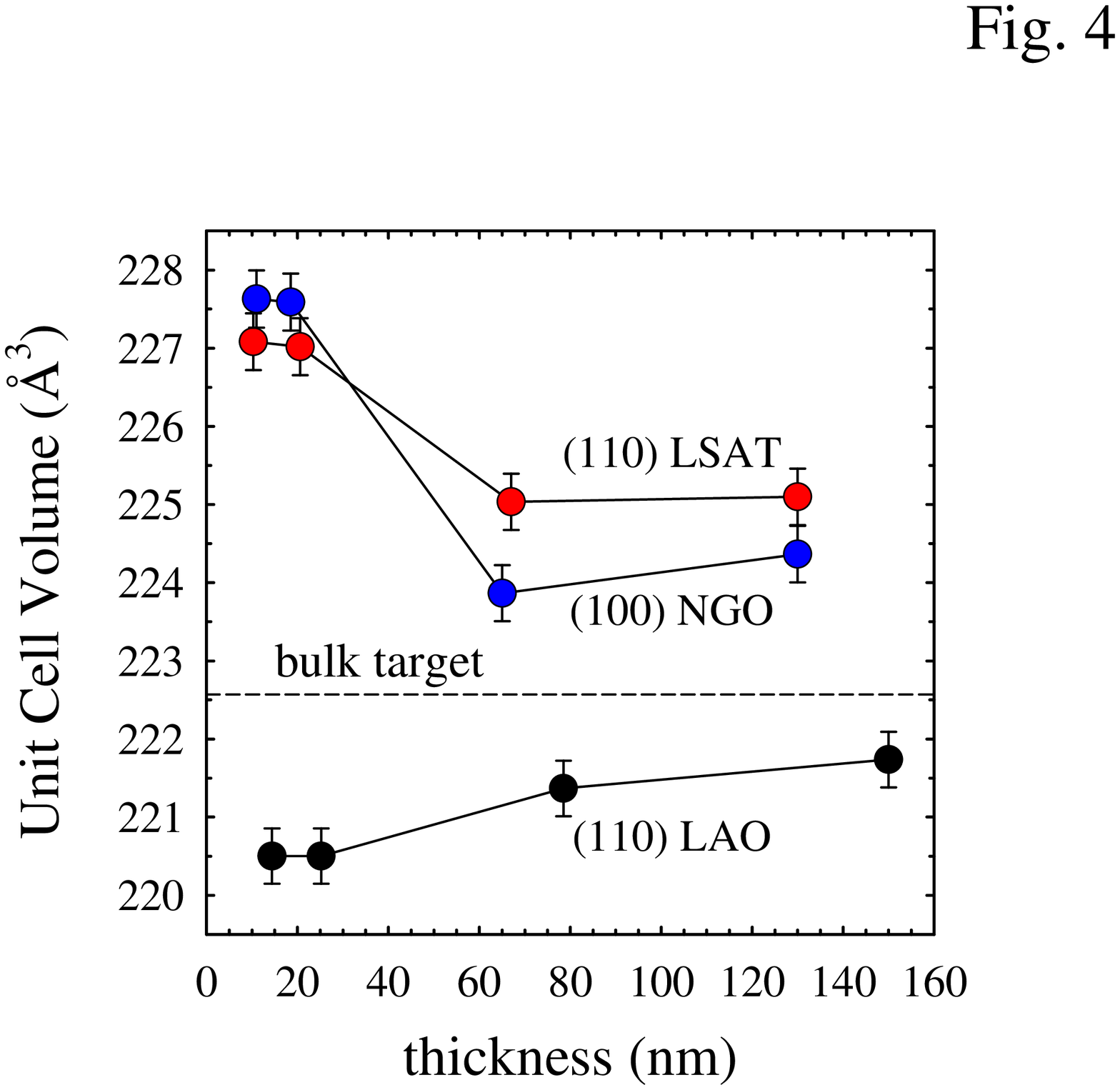}
\caption{(color online) Unit cell volume {\it vs} thickness for NSMO films on three substrates.}
 \label{CellVolume}
\end{figure}

For LAO films the result is an expansion along [010] and contraction along the film normal ([100]), whereas for NGO films the
$b$-axes contract and the $a$-axes expand.  For both substrates the thickest film is tetragonal.  The films on LSAT
exhibit more modest contractions along both [010] and [100] with increasing thickness, maintaining tetragonality.
In spite of these differences in behavior for the NGO and LSAT films under tensile strain, their unit cell volumes
(Fig.~\ref{CellVolume}) show very similar decreases with increasing thickness.

\begin{figure}
\includegraphics[width=4in,clip]{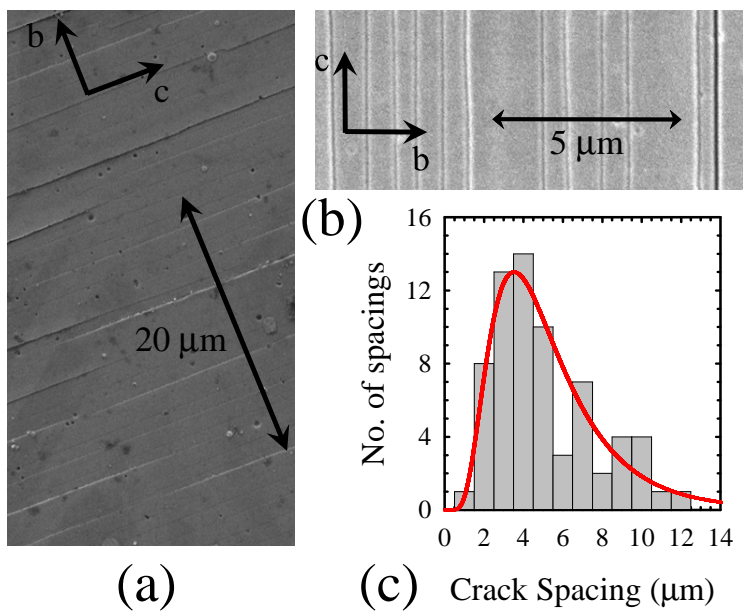}
\caption{(color online) SEM images of (a) 67-nm film on (110) LSAT, (b) 130-nm film on (100) NGO, and
(c) distribution of crack spacings for the NGO film. The solid curve is a log-normal distribution with median $=3.5\mu{\rm m}$
and geometric std. dev. $=1.26\mu{\rm m}$.}
 \label{SEM}
\end{figure}

Scanning electron micrographs (Fig.~\ref{SEM}) demonstrate that the relaxation of tensile strain along [010]
for the films on NGO and LSAT is accommodated by the formation of unidirectional crack arrays running along the film [001] direction.
The spacing of these cracks, determined from analyses of larger-area images from
films on each of the substrates, approximately describes a log-normal distribution (shown in Fig.~\ref{SEM}~(c)for the 65-nm NGO film), with
median values of $\sim 1-2\mu{\rm m}$ and $\sim 3-4\mu{\rm m}$ for LSAT and NGO films, respectively.  These distributions did not change
appreciably with thickness for either substrate.  The width of the cracks themselves varies within a given film (particularly
evident in Fig.~\ref{SEM} (a) for the NGO film), and the mean crack width is greater in the thicker films.  Though transverse-sectional
microscopy was not pursued, as we discuss further below the transport data implies that the cracks do not penetrate through to the substrate.

\begin{figure}
\includegraphics[width=3.375in,clip]{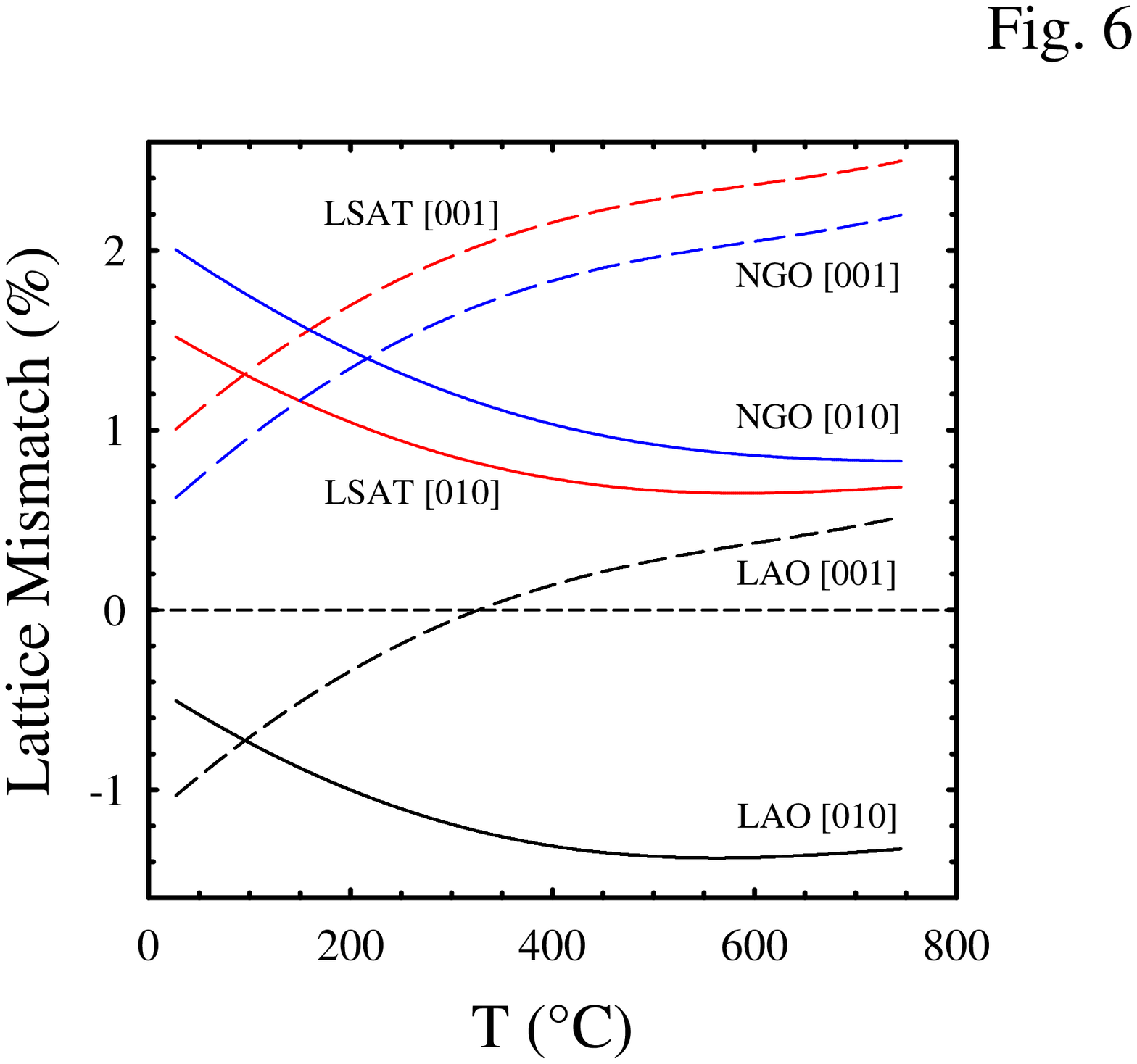}
\caption{(color online) Lattice mismatch {\it vs} temperature for the three substrates along the film [010] and [001] directions. Positive values
correspond to a substrate lattice constant larger than that of the film.}
 \label{Misfit}
\end{figure}

Similar crack arrays were observed previously\cite{Olsson} for [110]-oriented YBa$_2$Cu$_3$O$_{7-\delta}$ and PrBa$_2$Cu$_3$O$_{7-\delta}$
films grown on [110] SrTiO$_3$, where they were attributed to anisotropic thermal expansion mismatch between substrate and film
upon cooling from the growth temperature. The same mechanism appears applicable to the present oxide film-substrate systems since, as noted above,
NSMO has thermal expansion coefficients of opposite sign: positive along [010] and negative along [001].
The lattice mismatch, $(a_{sub}-a_{NSMO})/a_{sub}$ ($a_{sub}$ and $a_{NSMO}$ are the substrate and bulk target lattice constants, respectively)
is shown as a function of temperature in Fig.~\ref{Misfit} along the [010] and [001] film directions for each of the three substrates.
These curves were computed using published thermal expansion data for the substrates.\cite{LAOLSATExp,NGOExpan}  The target lattice
constants were measured up to 200$^{\circ}$C and their temperature
dependencies found to match well those of Tobe {\it et al.} (Ref.~\onlinecite{Tobe}) measured over a broader temperature range for
compounds with a slightly different stoichiometry; the target data were then extended to higher temperature using the suitably scaled expansion data.

At the growth temperature (750$^{\circ}$C) the tensile mismatch for LSAT and NGO is greatest along the film [001] direction.
Upon cooling, the mismatch along [001] decreases since the c-axis expands, while that along [010] increases. For the films on
NGO the [010] mismatch approaches 2\% at room temperature, the same amount by which the $b$-axis lattice parameter decreases abruptly
with increasing thickness.  Although the calculated compressive mismatch along [010] for films on LAO is only 0.5\% at room temperature, the compressed
lattice is only stable at low thicknesses.  Evidently there is a comparable critical thickness for NSMO above which both compressed and
expanded lattices are relaxed.

\begin{figure}
\includegraphics[width=3.375in,clip]{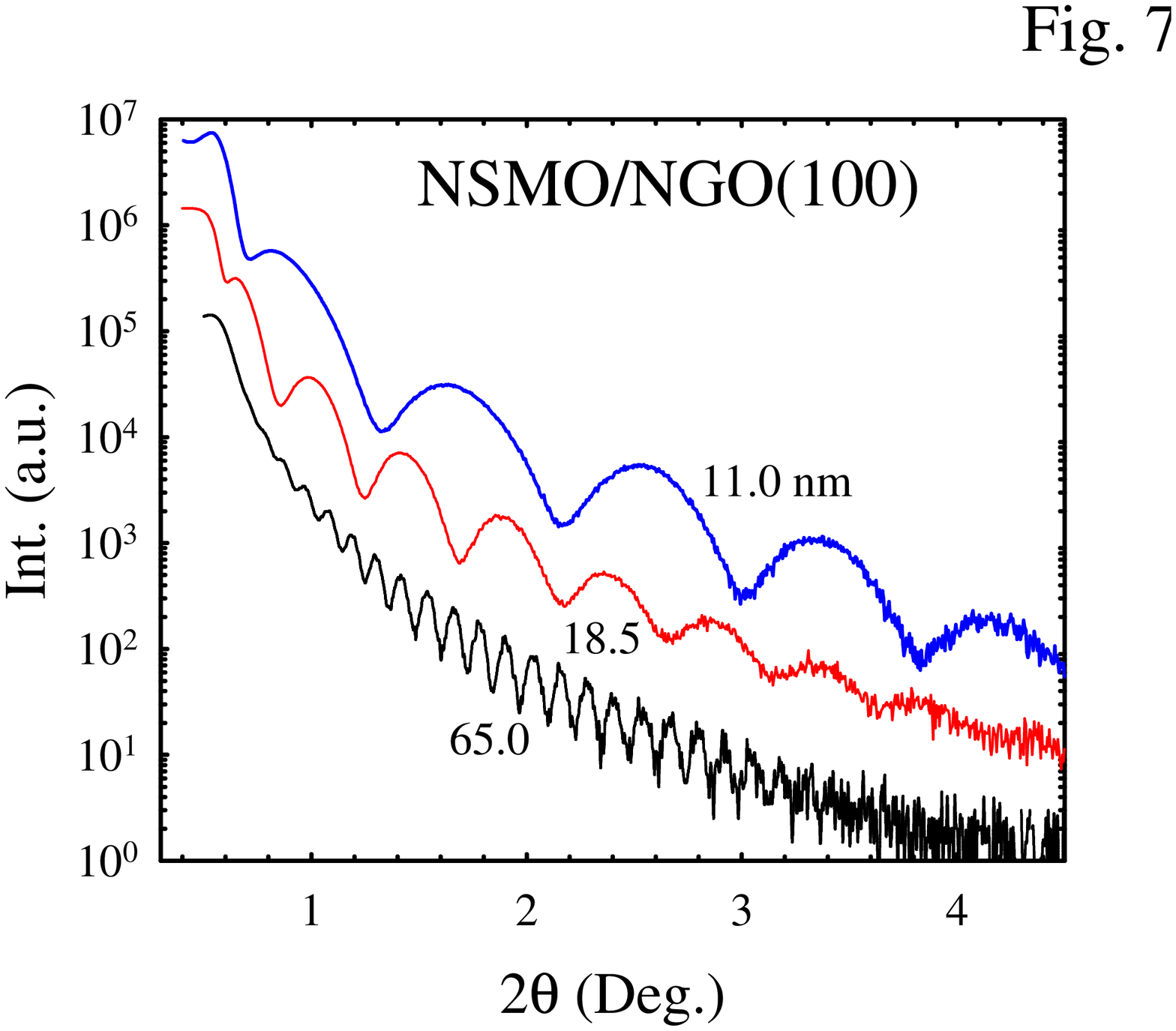}
\caption{(color online)Fig.(a) X -ray reflectivity for three films deposited on (100) NGO.}
 \label{XRR}
\end{figure}

In spite of the linear crack arrays that develop in the thicker films under tensile strain, all of the films remain smooth as
indicated by well-defined Kiessig oscillations seen in x-ray reflectivity measurements over an extended angular range
(Fig.~\ref{XRR}). Reflectivity simulations\cite{Spirkle} imply a film surface roughness with variance $\sigma\sim 0.4$nm for the
thinnest films, comparable to the perovskite unit cell dimension, with a modest increase for thicker films.

\subsection{\label{Transport} Transport Properties}

\begin{figure}
\includegraphics[width=3.375in,clip]{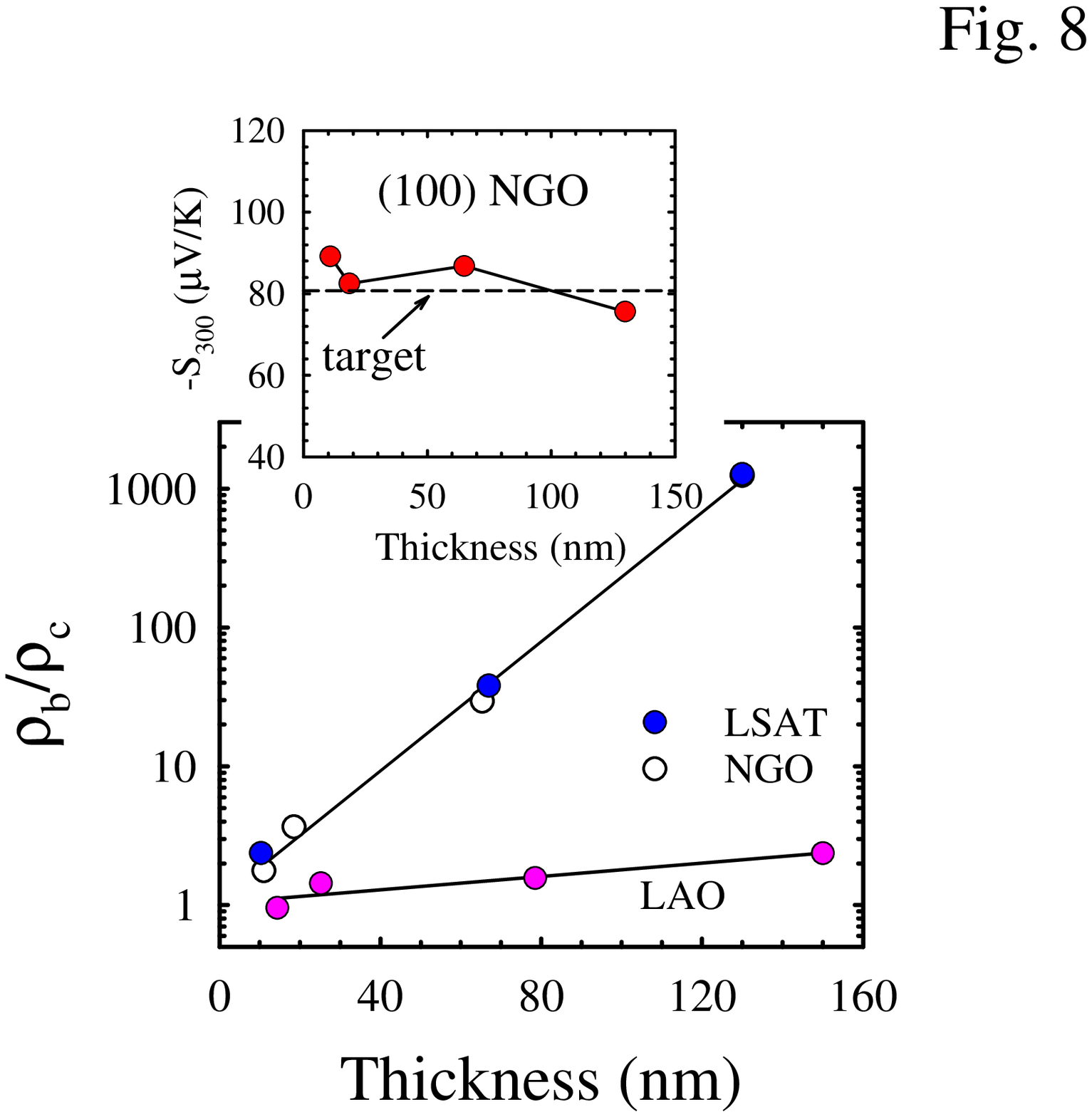}
\caption{(color online)Room-temperature in plane resistivity ratio, $\rho_b/\rho_c$, {\it vs.}
thickness for films grown on three substrates.}
 \label{RvsThickness}
\end{figure}

The dc electrical resistivity was measured for each film along the [010] and [001] directions as a function of temperature for $T\leq 325$~K.
Film resistances exceeding 1 G$\Omega$ prevented measurements below $\sim 50$~K.  The room-temperature resistivity anisotropy, $\rho_b/\rho_c$,
is found to increase approximately exponentially with increasing film thickness for the LSAT and NGO films (Fig.~\ref{RvsThickness}),
reflecting principally an increase along [010] due to cracking.  For the thickest films $\rho_b/\rho_c\simeq 1000$.  The corresponding ratio for
the LAO films also increases with thickness, reaching $\rho_b/\rho_c\simeq 2.4$ for the 150-nm film.  As this thickest film is tetragonal with
unit cell volume and $T_N$ closest to the bulk target (Fig.~\ref{LatticeConstants}), we take the latter value to be representative of the
intrinsic anisotropy of the NSMO compound.

To investigate the possibility that variations in oxygen content with thickness might contribute to the changes in
resistivity, the room-temperature thermopower (TEP) was measured for all films.  The TEP is a sensitive measure of
Mn valence that is largely independent of cation in this region of the manganite
phase diagram\cite{Raveau,CohnCLMO}. The value measured for the target was $-80\mu{\rm V/K}$.  The thermopowers for the films
did not vary by more than $\sim 10\%$ for all thicknesses, typically falling in the range $-(80-100)\mu{\rm V/K}$, as shown for the NGO films in
the inset of Fig.~\ref{RvsThickness}.  Thus significant and systematic variations in the oxygen content of the films with thickness are ruled out.

\begin{figure}
\includegraphics[width=3.375in,clip]{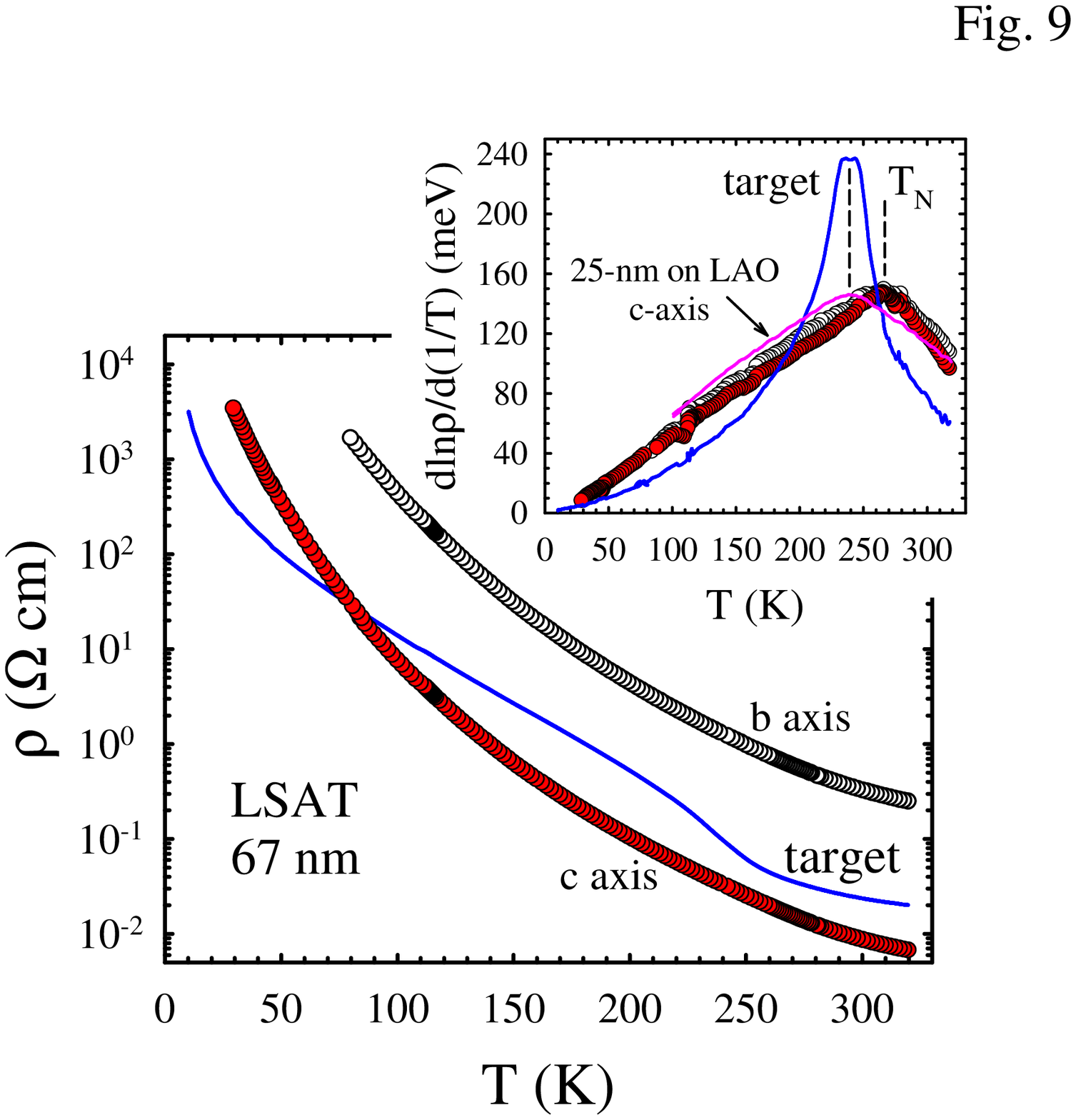}
\caption{(color online) $\rho(T)$ along the $b$ and $c$ axes of the 67-nm film grown on (110) LSAT.  The solid curve represents the
resistivity of the bulk, polycrystalline target.  The inset shows $d\ln\rho(T)/d(1/T)$ {\it vs.} $T$ for the same specimens and for
the 25-nm LAO film along the $c$ axis.}
 \label{NSMOrhovsT}
\end{figure}

Fig.~\ref{NSMOrhovsT} shows the $T$ dependence of the in-plane resistivities for the 67-nm film grown on (110) LSAT along with that of the target.
Interestingly, in spite of the substantial resistive anisotropy ($\rho_b/\rho_c\simeq 35$ for this film), the $T$ dependencies are essentially
identical along the [010] and [001] directions (inset, Fig.~\ref{NSMOrhovsT}).  This implies that the cracks do not penetrate into the substrate,
since a stronger temperature dependence would be expected associated with thermally activated tunnel conduction as observed
in other cracked manganite films.\cite{Fisher}

The antiferromagnetic transition temperatures, $T_N$, determined from maxima in $d\ln\rho(T)/d(1/T)$ {\it vs.} $T$
(inset, Fig.~\ref{NSMOrhovsT}), are $265\pm 3$~K for
LSAT and NGO films, and $T_N=240\pm 3$~K, the same as the bulk target, for films grown on LAO.
In G-type AF's like Ca$_{1-y}$Sr$_y$MnO$_3$, expansion of the lattice due to Sr substitution\cite{Chmaissem} enhances AF superexchange interactions
(and $T_N$) due to an increase in the Mn-O-Mn bond angle.  Stability of the C-type AF ordered state is increased by structural modifications tending
to isolate the FM one-dimensional chains.\cite{CAFStability}  In the present films
it seems likely that the significant enhancement of $T_N$ for films with expanded unit cells is
principally due to the expansion along [001].  Though enhanced superexchange, e.g.
due to an increase in the Mn-O-Mn bond angles along [100] and [010] is possible, the observation that $T_N$
for these films is independent of thickness, in spite of the very different $b$-axis lengths due to cracking in the thicker films,
argues against its prominent role in determining the increase in $T_N$.  More likely is that the expansion along [001]
actually enhances the double-exchange coupling along [001] in this compound, thereby increasing the stability of the orbital
and C-type N\'eel ordering.\cite{CAFStability}  We note that the $c$-axis length in the bulk target is
$\simeq 1.2$~\% smaller than that of a $x=0.75$ compound\cite{Kajimoto} with $T_N\simeq 300$~K.  Thus the observed increase in $T_N$
for the films under tension is consistent with existing data on the structure and phase behavior in this regime of composition,
where $T_N$ is plausibly controlled principally by the $c$-axis length.

\section{\label{sec:CONCL} CONCLUSION}

Epitaxial $a$-axis-oriented films of Nd$_{0.2}$Sr$_{0.8}$MnO$_3$ have been grown under both compressive and tensile strain.
The intrinsic resistivity anisotropy of the material, inferred from transport measurements on a 150-nm thick, compressively
strained film, is $\rho_b/\rho_c=2.4$.  Uniaxial crack arrays oriented along the film [001] direction develop in films
under tensile strain with thickness $\gtrsim 65 {\rm nm}$ due to anisotropic thermal expansion mismatch upon cooling from the growth temperature.
Typical crack spacings are a few $\mu{\rm m}$.  The resistivity anisotropy measured for the cracked films exceeds $10^3$.  Identical temperature
dependencies of the resistivity along and transverse to the cracks implies that not all penetrate through to the substrate,
and thus the large anisotropy is attributed to a thin and meandering conduction path for transport along [010].  The increased
N\'eel temperature for films under tensile strain, from 240~K for the bulk target to 265~K independent of thickness, suggests that
enhanced FM exchange stabilizes the orbital and spin order of the C-type AF state, and that their stability in this region of the phase
diagram is largely controlled by the $c$-axis length.

\bigskip
\textbf{ACKNOWLEDGMENTS}
\bigskip

We thank Mr. Alsayegh Husain for technical assistance with SEM scanning. This material is based upon work supported by the
National Science Foundation under grants DMR-0072276 (Univ. Miami) and DMR-0504769 (Montana State Univ.),
the Research Corporation (Univ. Miami), and the U.S. DOE Office of Basic Energy Sciences (Montana State Univ., Grant No. DE-FG-06ER46269).

\noindent
$^{\dagger}$ present address: 201-1650 Pembina Hwy., Winnipeg, MB R3T2G3 CA

\newpage
%

\end{document}